# Social Network Mining (SNM): A Definition of Relation between The Resources and SNA

Mahyuddin K. M. Nasution

*Information Technology Department, Fakultas Ilmu Komputer dan Teknologi Informasi, Universitas Sumatera Utara,
Padang Bulan, Medan 20155 Sumatera Utara, Indonesia
E-mail: mahyuddin@usu.ac.id*

*Abstract*— Social Network Mining (SNM) has become one of the main themes in big data agenda. As a resultant network, we can extract social network from different sources of information, but the information sources were growing dynamically require a flexible approach. To determine the appropriate approach needs the data engineering in order to get the behavior associated with the data. Each social network has the resources and the information source, but the relationship between resources and information sources requires explanation. This paper aimed to address the behavior of the resource as a part of social network analysis (SNA) in the growth of social networks by using the statistical calculations to explain the evolutionary mechanisms. To represent the analysis unit of the SNA, this paper only considers the degree of a vertex, where it is the core of all the analysis in the SNA and it is basic for defining the relation between resources and SNA in SNM. There is a strong effect on the growth of the resources of social networks. In total, the behavior of resources has positive effects. Thus, different information sources behave similarly and have relations with SNA.

*Keywords*— extraction; vertex; edge; actor; multiple regression; α-Cronbach

## I. INTRODUCTION

Tendency of increasing the aspects of life involving social networks is not only aligned with the popularity of social networking sites such as Facebook, Twitter, VKontakte, QZone, Odnoklassniki, etc., but also as a result of the growth of information on the Web (as information sources) [1], [2]. In this case, the computer network as the social network [3] implies that the Web as social media containing the big data [4] to be a big picture of the real world, and in the structure we can express it through the social network [5], [6]. Currently, one of the results of social network is a resultant from the methods of social network extraction, whereby actors/vertices, relationship/edges, and web/documents are resources of social network [6], [7]. Social network as information about the behaviour of social actors to be important in decision making such as the importance of information source like Web.

In the social sciences, Social Network Analysis (SNA) developed from a conjecture of anthropologist's observations [8] about relation in the face-to-face groups [9] and it based on mathematical graph theory [10]. Unlike generating a conventional social network, on the one hand, extracting social network from Web dealing with everything that changed dynamically [11], i.e. enormous amount of information of the social actors and the clues about relations among them. On the other hand, every time we did not just find the new web pages, but also the presence of new actors. This led to the extraction of social networks be more complex and involves data increasingly large, while extracting the social network depends heavily on the limited services of tools such as search engine [12]. Therefore, extracting the social networks from information sources like the Web relatively just based on samples [7], and it is useful to learn the behaviour of the social network. That is, Social Network Mining (SNM) to provide a means of discovering the behaviour of either the information sources or the resource, SNM is not same as the SNA, but to overcome a bit of information connectedness between them. Thus, this paper aimed to express the relation among vertices, edges, and Web in the growth of social networks.

This paper based on the conceptual bridge about SNM and SNA. Therefore, this paper consists of four sections: In Section II we present what already known as the material and method. While in Section III is an attempt to present information through cognitive structures and outlines them as a presentation for interpreting them to new things in results and discussion. The last section in the form of summaries or the conclusion, whereby based on it we reveal an issue for future work.

## II. MATERIAL AND METHOD

Social Network Mining (SNM) have been declared as social network data mining [11] is to refer to the definition of the data mining: a process for discovering useful information automatically from the Web such as the large data repositories. The SNM techniques are deployed to explore large structure in order to formulate rules that can adopt useful patterns where they may have not unknown [7]. To obtain the social structures from the unstructured information, where one of them is a social network, is by extracting the information (social network) from the Web [12]. In other words, SNM or simply the network mining [13] fully become a part of knowledge discovery in a social structure, which is the pre-processing that transforms the raw input data into a social network. For subsequent analysis, the process that converts the structured data into useful information [14], and a post-processing for ensuring that valid and useful results for the decision support system [15].

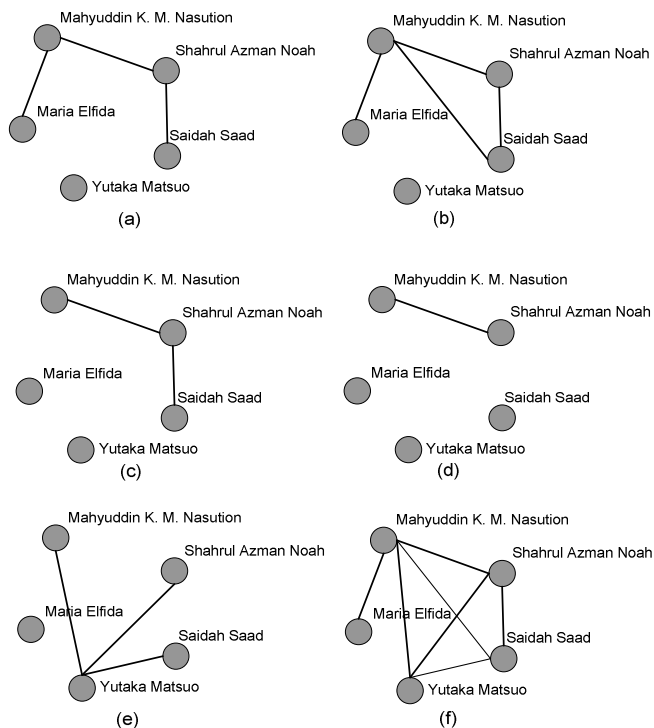

Fig. 1 Some of the relationships among five social actors

The term "network" does not have the same meaning in different fields [16]. In a dynamic context, social network is a collection of information about ties between all pairs of social actors where a tie connects a pair of actors by one or more relations, thus the network data fully give a complete picture about relations in the population [17]. The Web has been becoming the largest text database contained information about social actors, but the web characterized unstructured, insufficient and incomplete information. Therefore, SNM involves the information extraction, i.e. a systematic approach for the study of indiscernibility of social actors. Thereof, the social network extraction requires a semantic technology such as the use of co-occurrence to collect the clues of the relation between actors [12].

Naturally, a social network can be modeled by a graph $G<V,E>$ whereby either SNA or SNM use the similar approach to visualize a network that consists of $V$ as a set of vertices and $E$ as set of edges [18]. In statistical literature, the researchers have the general model of a social network as a Cartesian product of $n$ actors with generating their relations [19], [20]. Cartesian product can be represented as an $n \times n$ matrix M [21]: An edge $e_{kl} = m_{kl}$ in M is 1 for all $e_{kl}$ in E, if a pair of vertices is adjacent of $v_k$ in V and $v_l$ in V, $e_{kl} = 0$ otherwise. For example, Fig. 1 possess matrices: $M_{5(a)}$, $M_{5(b)}$, $M_{5(c)}$, $M_{5(d)}$, $M_{5(e)}$ and $M_{5(f)}$ as follows [22].

|  | a. | b. | c. | d. | e. |
|---|---|---|---|---|---|
| a. Mahyuddin K. M. Nasution | 1 | 1 | 0 | 1 | 0 |
| b. Shahrul Azman Noah | 1 | 1 | 1 | 0 | 0 |
| c. Saidah Saad | 0 | 1 | 1 | 0 | 0 |
| d. Maria Elfida | 1 | 0 | 0 | 1 | 0 |
| e. Yutaka Matsuo | 0 | 0 | 0 | 0 | 1 |

|  | a. | b. | c. | d. | e. |
|---|---|---|---|---|---|
| a. Mahyuddin K. M. Nasution | 1 | 1 | 1 | 1 | 0 |
| b. Shahrul Azman Noah | 1 | 1 | 1 | 0 | 0 |
| c. Saidah Saad | 1 | 1 | 1 | 0 | 0 |
| d. Maria Elfida | 1 | 0 | 0 | 1 | 0 |
| e. Yutaka Matsuo | 0 | 0 | 0 | 0 | 1 |

|  | a. | b. | c. | d. | e. |
|---|---|---|---|---|---|
| a. Mahyuddin K. M. Nasution | 1 | 1 | 0 | 0 | 0 |
| b. Shahrul Azman Noah | 1 | 1 | 1 | 0 | 0 |
| c. Saidah Saad | 0 | 1 | 1 | 0 | 0 |
| d. Maria Elfida | 0 | 0 | 0 | 1 | 0 |
| e. Yutaka Matsuo | 0 | 0 | 0 | 0 | 1 |

|  | a. | b. | c. | d. | e. |
|---|---|---|---|---|---|
| a. Mahyuddin K. M. Nasution | 0 | 1 | 0 | 0 | 0 |
| b. Shahrul Azman Noah | 1 | 0 | 0 | 0 | 0 |
| c. Saidah Saad | 0 | 0 | 0 | 0 | 0 |
| d. Maria Elfida | 0 | 0 | 0 | 0 | 0 |
| e. Yutaka Matsuo | 0 | 0 | 0 | 0 | 0 |

|  | a. | b. | c. | d. | e. |
|---|---|---|---|---|---|
| a. Mahyuddin K. M. Nasution | 0 | 0 | 0 | 0 | 1 |
| b. Shahrul Azman Noah | 0 | 0 | 0 | 0 | 1 |
| c. Saidah Saad | 0 | 0 | 0 | 0 | 1 |
| d. Maria Elfida | 0 | 0 | 0 | 0 | 0 |
| e. Yutaka Matsuo | 1 | 1 | 1 | 0 | 0 |

|  | a. | b. | c. | d. | e. |
|---|---|---|---|---|---|
| a. Mahyuddin K. M. Nasution | 1 | 1 | 1 | 1 | 1 |
| b. Shahrul Azman Noah | 1 | 1 | 1 | 0 | 1 |
| c. Saidah Saad | 1 | 1 | 1 | 0 | 1 |
| d. Maria Elfida | 1 | 0 | 0 | 1 | 0 |
| e. Yutaka Matsuo | 1 | 1 | 1 | 0 | 1 |

Fig. 1(a) and matrix $M_{5(a)}$ explain the relationship among five social actors in their activities their separately. While Fig. 1(b) and matrix $M_{5(b)}$ reveal that among five social actors there is at least one paper together. Fig. 1(c) and matrix $M_{5(c)}$ show author-relationship for five social actors, with which Fig. 1(d) and matrix $M_{5(d)}$ mark only the relationship between a supervisor and a student. Fig. 1(e) and matrix $M_{5(e)}$ reveal that some of the authors conduct citation against one another author. The last, Fig. 1(f) and matrix $M_{5(f)}$ are the summaries of possible relationships between five social actors.

In pre-processing as a part of the overall process of SNM, the extracted social network (resultant) that is represented as $SN = <V,E,A,R,\gamma_1,\gamma_2>$ with the conditions as follow [22]:

(a) $\gamma_1(1:1) : A \rightarrow V$, $A = \{a_i|i=1,\ldots,n\}$ is a set of social actors, and
(b) $\gamma_2 : R \rightarrow E$ or $e_j = \gamma_2(r_s(a_k,a_l))$ where $R = \{r_s|s=1,\ldots,p\}$ is a set of relations, and $a_k,a_l$ in $A$.

Therefore, social networks can grow in a comparison pattern of star-graph ($n$-1) and the complete-graph ($n(n-1)/2$). In other words, the first condition states that any actor's name refers to vertex individually, although the possibility of the actor literally represented by different name text, so there are several approaches that each actor represented by a text name, for example by adding keyword [23]. Next, unlike the classic relationship between social actors, a relationship based on the information source generally expressed in the strength relation. Therefore, the second condition defines that a relationship between two actors consists of a set of relations [22]. Upon consideration of the provisions relating to social networks, the extraction method exploits search engine for obtaining documents related to the participating actors in occurrence and co-occurrence, subsequently calculate the strength relation [19], and we call it as superficial method [12].

Today, in daily life, anywhere, the information about social actors and their relations are so important. For any application, every social network, including the extracted social network (the resultant of extraction methods) [24], can be analysed to obtain the behaviour of a social community, such as the size, density, degree, reachability, distance, diameter, geodesic distance, etc., they have been disclosed in the literature of SNA [25]. SNA is a study of social networks for interpreting the structural relationships between actors, or in traditional SNA is descriptive. Therefore, kind of SNA requires more sophisticated measurements, i.e. a study of social network for discovering knowledge from information sources [26], we call it as SNM or it is predictive [27], by using automatic learning and analysing the content of information, or by involving the clustering techniques to identify the relevant content.

SNA, in general, involves a part of social network resources (vertices and edges) in a unit analysis [15], but SNM involves all resources (vertices, edges, and information sources) and plays a role as a bridge between the social network resources and SNA. Information sources are like Web, corpus, or documents. As different units: vertices and edges have characteristics, respectively. Their characteristics not only determine their behaviour, but they determine behaviour of relation among them [28], [29]. Although the social network conceptually has designed to map actors relationship that can be observed, but in SNA is to mark patterns of ties between actors and to present a variety of social structure according to interests.

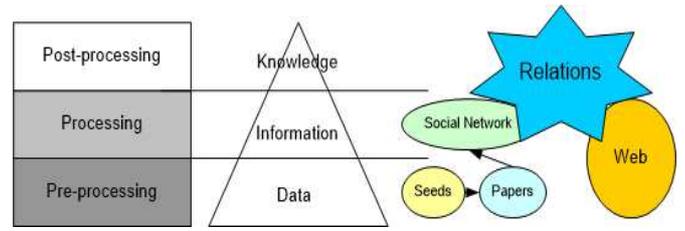

Fig. 2 An approach to defining the relation resources and SNA

For defining the relation between the resources: seeds, actors, vertices, edges, papers, web pages and degree of actors, based on Fig. 2, we conduct some computations as follows, where the degree of vertices as the basis of SNA [25]. First, we collect a set of actors as seeds (sample) for generating social networks [7], and we conduct a random test of sample against sorted names. Seeds and papers are data obtained during the pre-processing of the SNM. Second, we collect papers for each seed and we conduct extracting social networks from documents based on publication year. For each seed, we have a timeline of social network: The growth of the vertices and edges by a growing number of papers. Third, we collect the hit counts for patterns of actor name or the hit counts for actor names in a query [30], [31], [32]. We test the randomness of the sample to be able to represent the social communities using test runs. Firstly, we test for a sequence of the actor names based on academic level, and then we test the sequence of the number of papers, the hit counts with quotation marks, and the hit counts as validation that there is a relation between seeds and information source (papers and web pages) [33]. In this case, the hit counts as a representation of the big data that presented by any search engine [34], [35], [36]. Based on the principle and assumption about the sample: If the population can converge into a big data and then it can become sample [12], then a sample of the actor population is the seed, and a sample of the web pages population is an online database such as DBLP. In this case, an online database according to its size can refer to as a big data. Therefore, the proposed method is based on the principle of extraction can be used for evaluating the big data [7].

We use the multiple regression to generate the behaviour of relations between resources. In the multiple regression, the independent variables $x_i$, $i=1,\ldots,n$ and dependent variable $y$, the average of $y|x_i$ given by linear regression models, i.e.

$$\mu_{y|xi} = b_0 + \Sigma_{i=1\ldots n} b_i x_i \qquad (1)$$

and the estimation of responses obtained from the regression equation of a sample is

$$y = \beta_0 + \Sigma_{i=1\ldots n} b_i x_i. \qquad (2)$$

We calculate a total relation as follows

$$tr = \Sigma_{i=1\ldots n} \Pi_{j=1\ldots i} \beta_j \qquad (3)$$

where $\beta_1 = \Pi_{j=1} \beta_j$ means a direct effect and $\Pi_{j=2\ldots n} \beta_j$ means the indirect effect. For a sample of the quantities which consists of $k$ clusters may be measured by $Y = \Sigma x_i$, $i=1,\ldots,n$.

In order to reduce the constant of internal consistency of the sample behaviour in general, we use $\alpha$-Cronbach or

$$\alpha = (k/(k-1))(1 - \Sigma_{i=1...n}\, \sigma^2_{xi}/\sigma^2_Y) \qquad (4)$$

by which $\sigma^2_Y$ is the variance of the score total was observed while $\sigma^2_{xi}$ was the variance of $i$-component for sample $x_i$. Variance is calculated by using $\sigma^2_{xi} = 1/n\, \Sigma(x_i\text{-x})$, x is average of $x_i$. In general, rule of use $\alpha$ is to use the marker of

(a) $\alpha > 0.9$: The internal consistency of behaviour is very good,
(b) (b) $0.7 \leqslant \alpha \leqslant 0.9$: The internal consistency of behaviour is good,
(c) (c) $0.6 \leqslant \alpha \leqslant 0.7$: The internal consistency of behaviour is acceptable,
(d) (d) $0.5 \leqslant \alpha \leqslant 0.6$: The internal consistency of behaviour is poor, and
(e) (e) $\alpha < 0.5$: The internal consistency of behaviour is not accepted.

Consistency states that information maintained from time to time. In other words, the data that is processed can be replaced by similar data and generate the similar characteristics. For example, in similar format we can replace the random data with other random data, but cannot replace it with data is not random. Randomness be the characteristics of the data, and becoming behaviour for social actors associated with that data.

TABLE I
THE GROWTH OF NETWORKS OF TWO SEEDS

| Year | Actor: AAB | | | Actor: SAMN | | |
|---|---|---|---|---|---|---|
| | Paper | Vertice | Edge | Paper | Vertice | Edge |
| 1995 | | | | 1 | 2 | 1 |
| 1996 | | | | 1 | 2 | 1 |
| 1997 | | | | 1 | 2 | 1 |
| 1998 | | | | 2 | 2 | 1 |
| 1999 | | | | 3 | 2 | 1 |
| 2000 | | | | 4 | 3 | 2 |
| 2001 | 1 | 4 | 6 | 4 | 3 | 2 |
| 2002 | 2 | 4 | 6 | 5 | 5 | 5 |
| 2003 | 2 | 4 | 6 | 6 | 12 | 33 |
| 2004 | 2 | 4 | 6 | 8 | 13 | 37 |
| 2005 | 3 | 6 | 9 | 14 | 19 | 51 |
| 2006 | 3 | 6 | 9 | 15 | 20 | 51 |
| 2007 | 4 | 7 | 11 | 17 | 23 | 55 |
| 2008 | 22 | 15 | 30 | 18 | 24 | 56 |
| 2009 | 37 | 40 | 189 | 23 | 42 | 198 |
| 2010 | 42 | 53 | 219 | 32 | 57 | 238 |
| 2011 | 50 | 60 | 265 | 41 | 73 | 275 |
| 2012 | 59 | 68 | 295 | 47 | 76 | 281 |
| 2013 | 60 | 68 | 295 | 47 | 76 | 281 |
| 2014 | 64 | 72 | 302 | 48 | 80 | 304 |
| 2015 | 69 | 80 | 329 | 53 | 86 | 313 |

III. RESULT AND DISCUSSION

In this experiment, we have defined as many as 37 names of actors as the seeds to generate other actors and to build social networks. Actor names mentioned we collected from website of Faculty of Information Science & Technology, Universiti Kebangsaan Malaysia (http://www.ftsm.ukm.my/). Actors divided into 2 categories based on academic level (al), that is 13 Professors (pr) and 24 Associate Professors (ap) as shown in Fig. 3.

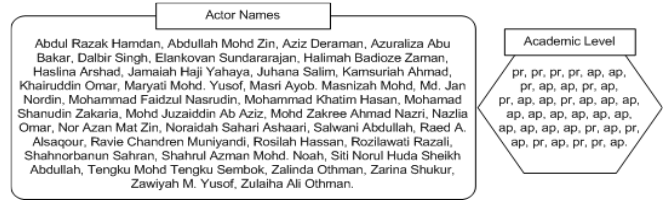

Fig. 3  List of seeds as a sample

Suppose taken a name of social actor as a seed, i.e. "Azuraliza Abu Bakar" (AAB). A number of papers ($n(p)$) written by the actor until 2015 can be obtained from any online databases, such as DBLP [6], i.e. $n(p) = 69$. The scientific papers that have been published every year from the first year until 2015 we structure into a database, where the enhancement about the number of actors and number of edges that form a social network based on seed are like in Table 1 (first rows). Every year for each seed, we can collect the new actors and relationships between them based on the concept of co-occurrence. We get first: $nes$ = the number of edges based on the number of actors where the seed as the centre of social networks (star graph) in a complete graph, and the next: $ner$ = number of edges generated (reality graph) in a complete graph. Comparison between $nes$ and $ner$ revealed the existence possibility of sub-networks as small groups build the new collaborations if the star graph and the reality graph are not overlapped. For example, until 2015 there were 80 vertices for the actor "Azuraliza Abu Bakar", with a vertex degree $n$-1 = 79, but in the social network, there are 313 edges: resulting in more than one vertex having a degree > 1 besides the centre vertex. So there are several leaves of star graph interconnected and then grew into a branch of leaves others. In this case, we generate relations between the papers, the social actors, and their relationships based on seed as follows:

- If start of reality/complete is 1, then number of authors for first paper is 2 actors;
- If start of reality/complete is less than 1, then number of authors for first paper is more than two actors;
- If start of reality/complete is 0, then number of authors for first paper is an actor (the paper was written independently).

Then, the value of reality/complete smaller than 1, indicating the growth of social networks and also the dissemination of knowledge continuous. Each actor as the seed has published a number of papers as the source of information whereby web pages as social media also can reflect the activities of seed. That is, some results (the hit counts) that returned by search engine to respond to submitting query either with quotation mark or not, i.e. ("hc" ("Azuraliza Abu Bakar") = 4,740 hits) and (hc (Azuraliza Abu Bakar) = 5,480 hits), respectively, showed the social indication of actor's name. Likewise, the social actor "Shahrul Azman Mohd Noah" (SAMN), see Table 1 - second column. Behaviour of growth for each social network based on seed have shown that there is a positive direct effect $\beta_{11}$ of papers toward edges, but some seeds generate the negative direct effects $\beta_{11}$, likewise the indirect effects $\beta_1\beta_{12}$, where the negative direct effect on seed means that influence seed to the growth of other actors is getting smaller. For all seeds of social networks, the influence of each factor on the growth of social networks, as shown in Fig. 3, can be considered as similar. Therefore, if the hit counts can represent a social actor and the papers of seed can generate a social network, then the hit counts also generate a social network with the same behaviour.

In examining the behaviour based on the data or for the data to describe the behaviour, the characteristics of the data must be disclosed. Therefore, the corresponding data expressed through the data engineering. To do the data engineering in accordance with the needs of research related to big data, we must first review the basic behaviour of the population, namely the randomness of the sample: a test used to see whether the samples was taken at random so that samples can be representative of the population. A sequence of the actor names in alphabetical order (Fig. 2) with 18 runs ($r = 18$) toward the academic levels (al): *pr* and *ap*. Distribution of samples *r* approaches a normal distribution of $Z$: $npr = 13$ and $nap = 24$. Hypotheses are

$H_0$: Order of *pr* and *ap* in a row is random.
$H_1$: Order of *pr* and *ap* in a row is not random.

The function of the hypothesis is as temporary answer to the issue and still as a presumption, because they still have to be verified. Thus, in addition, to push for the emergence of the concept of the relationship between the SNA and the SNM, but also as a framework to draw up conclusions on research on the social network.

TABLE II
INFORMATION FOR RANDOM TEST

| Measurement | al | $n(p)$ | "hc" | hc |
|---|---|---|---|---|
| *npr* | 13 | 19 | 19 | 19 |
| *nap* | 24 | 18 | 18 | 18 |
| run *r* | 18 | 21 | 17 | 19 |
| average $\mu_r$ | 17.86 | 19.49 | 19.49 | 19.49 |
| variance $\sigma_r$ | 2.73 | 2.99 | 2.99 | 2.99 |
| $Z_{count}$ | 0.05 | 0.51 | -0.83 | -0.16 |

In this case, the average is $\mu_r = 17.87$ and the variance is $\sigma_r = 2.73$. By using $\alpha = 0.05$, $Z_{count} = 0.05 < Z_{0.25} = 1.96$. So $H_0$ accepted, or row of the actors as the seeds (sample) is random with a confidence level is 95 percent. For case $n(p)$ as number of papers in DBLP, we have $Z_{count} = 0.51 < Z_{0.25} = 1.96$ whereby $H_0$ accepted. While for "hc" and hc, we have calculated that $Z_{count}$ of "hc" and hc are negative, but $H_0$ accepted whereby $Z_{count} = -0.83 > Z_{0.25} = -1.96$ and $Z_{count} = -0.16 > Z_{0.25} = -1.96$, see Table 2. Thus, the sample based on all factors are the random.

In context about exploring the behaviour of growth of the social network, the prediction models of resources take heavy a position as a conduit of information about the dynamism of social network as part of social network mining. The multiple regressions are one of the methods to determine the causal relationships between factors as resources of the social network. First, we calculate it based on the accumulation

("1"): 1,1,1,1,1,2,2,3,3,4,5,8,9,11,12,19,20,23,25,28,30, 35,37,37,37,37,

TABLE III
FOUR FACTORS OF SOCIAL NETWORK IN TIMELINE

| Year | Resources | | | |
|---|---|---|---|---|
| | Seed | Paper | Vertices | Edges |
| 1990 | 1 | 1 | 2 | 1 |
| 1991 | 1 | 1 | 2 | 1 |
| 1992 | 1 | 1 | 2 | 1 |
| 1993 | 1 | 1 | 2 | 1 |
| 1994 | 1 | 1 | 2 | 1 |
| 1995 | 2 | 2 | 4 | 2 |
| 1996 | 2 | 3 | 6 | 5 |
| 1997 | 3 | 4 | 9 | 8 |
| 1998 | 3 | 5 | 9 | 8 |
| 1999 | 4 | 8 | 12 | 11 |
| 2000 | 5 | 12 | 18 | 25 |
| 2001 | 8 | 15 | 27 | 37 |
| 2002 | 9 | 20 | 33 | 48 |
| 2003 | 11 | 29 | 53 | 109 |
| 2004 | 12 | 35 | 59 | 123 |
| 2005 | 19 | 50 | 86 | 221 |
| 2006 | 20 | 63 | 99 | 245 |
| 2007 | 23 | 84 | 123 | 310 |
| 2008 | 25 | 107 | 139 | 353 |
| 2009 | 28 | 171 | 204 | 606 |
| 2010 | 30 | 223 | 251 | 734 |
| 2011 | 35 | 346 | 381 | 1107 |
| 2012 | 37 | 396 | 425 | 1249 |
| 2013 | 37 | 446 | 489 | 1440 |
| 2014 | 37 | 509 | 536 | 1588 |
| 2015 | 37 | 554 | 582 | 1714 |

While the second based on the accretion value

("2"): 1,0,0,0,0,1,0,1,0,1,1,3,1,2,1,7,1,3,2,3,2,5,2,0,0,0.

The first actor and the last actor appeared from 1990 to 2012, and the data in the range of 1990 to 2015, then model of r elation between seeds ($x_1$), papers ($x_2$), vertices ($x_3$), and edges ($y$) as follows: The dependent variable $x_2$ and independent variable $x_1$,

("1") $x_2 = 11.39 - 53.09 x_1$, (5)

("2") $x_2 = 6.88 + 11.52 x_1$; (6)

The dependent variable $x_3$ and independent variables $x_1$ and $x_2$,

("1") $x_3 = 0.93 + 1.78 x_1 - 0.11 x_2$, (7)

("2") $x_3 = 0.95 + 1.76 x_1 - 0.35 x_2$; (8)

The dependent variable $y$ and independent variables $x_1$, $x_2$, and $x_3$,

("1") $y = 1.54 + 1.49x_1 + 0.65x_2 - 13.91x_3$, (9)

("2") $y = 2.73 + 0.29x_1 + 1.87x_2 - 4.19x_3$. (10)

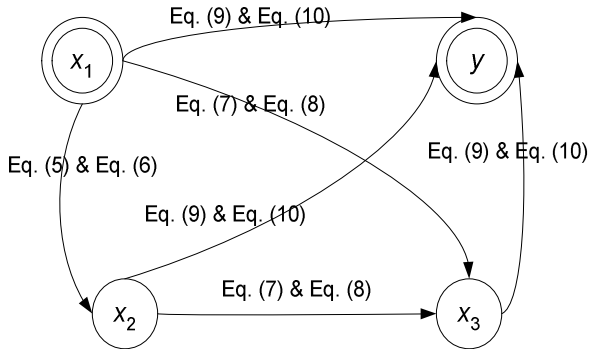

Fig. 4 Network of direct and indirect effects for four factors (resources)

From Equations (5), (7), and (9) we have the direct effect of the seeds toward edges is 1.49, while the indirect effect is $x_1x_2 + x_1x_3 + x_1x_2x_3 = (-53.09)(0.65) + (1.78)(-13.91) + (-53.09)(-0.11)(-13.91) = -34.77 - 24.79 - 82.82 = -142.39$ such that effect total is -140.91. From Equations (2), (4), and (6), the direct effect of the seeds toward edges is 0.29, while the indirect effect is $x_1x_2 + x_1x_3 + x_1x_2x_3 = (11.52)(1.87) + (1.76)(-4.19) + (11.52)(-0.35)(-4.19) = 21.56 - 7.38 + 16.65 = 30.83$ such that effect total is 31.12. Therefore, all the seeds individually have an impact on the growth of social networks directly and indirectly.

TABLE IV
β FOR EQUATION (2) FOR DATA IN TABLE 3

| 10 years | | Seeds | Papers | Vertices |
|---|---|---|---|---|
| Paper | 2.0826 | -1.2569 | | |
| Vertices | 0.3344 | 2.6981 | -1.0292 | |
| Edges | 1.4961 | 0.1555 | -2.0171 | -0.1680 |
| 20 years | | Seeds | Papers | Vertices |
| Paper | 4.6180 | -10.6814 | | |
| Vertices | 0.7859 | 2.5019 | -1.9328 | |
| Edges | 4.4244 | -0.2673 | -8.9701 | -2.8325 |
| 26 years | | Seeds | Papers | Vertices |
| Paper | 11.3962 | -53.0889 | | |
| Vertices | 0.9262 | 1.7828 | -0.1122 | |
| Edges | 1.5368 | 1.4867 | 0.6549 | -13.9110 |

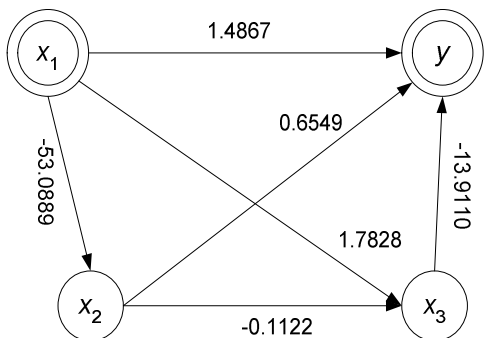

Fig. 5 Network of direct and indirect effects for resources

Based on Equations (2), (3) and Table 4 we have direct effect for 10 years is 0.1555 and indirect effect for 10 years (1990-1999) is (-1.2569)(-1.0292)(-0.1680) + (2.6981)(-0.1680) + (-2.0171)(-0.1680) = -0.33173, and effect total for 10 years is -0.1762. Similarly, we obtain the direct effect for 20 years is -0.2673 and indirect effect for 20 years (1990-2009) is -40.1558 with which the effect total for 20 years is -40.4231. Whereas for years 1990-2015, the direct effect = 1.4867 and the indirect effect = -116.773 and $tr$ = -115.286, see Fig. 5. In general, the effect of seeds on the growth of papers is positive, while the effect of seeds on the growth of other factors is a negative or positive. Thus, the effect total of seeds on growth social networks increasingly large but is negative. This means that the growth does not depend on the seeds only, but it depends on the amount of papers and the number of new actors.

As one of the factors, the scientific papers of actors (act as seeds) have a role in the growth of vertices and edges in social networks so that each social network has its own behaviour. At the time of conducting experiments, we have 37 $n(p)$, 37 "hc", and 37 hc as three collections of factors. Thus $k=3$. For 37 seeds, we have $\Sigma n(p) = 781$, $\Sigma"hc" = 540692$, and $\Sigma hc = 1468210$, and then $\Sigma_{i=1...37}\sigma^2_{xi} = 11978412919$ and $\sigma^2_Y = 20892233528$. Therefore, three factors are reliable for representing each other because α-Cronbach is 0.64 (acceptable). In this case, the median of $n(p)$, "hc" and hc individually are 14, 2800, and 8520. Thus we have number of data greater than median for $n(p)$, "hc", and hc are 19, number of data less than median for $n(p)$, "hc", and hc are 18, while based on the sequence of actor names we have transitions 21 for $n(p)$, 17 for "hc", and 19 for hc. Thus $Z_{0.25}=-1.96 < Z_{count}=0.51;-0.83;-0.20 < Z_{0.25}=1.96$ for $n(p)$, "hc" and hc, or behaviour of sample is random for three factors.

In SNA, the degree of a vertex is major categories in determining the role of an actor in community based on social networks: One of them is as central of research groups. Thus, each seed in social networks have degree greater than 1, and each actor connects with others by more than one edge. Therefore, it is possible, an actor as the seed first build a social network in form of star-graph or as the centre of other actors, but later became a centre of the research group. The behaviour of resources in a social network can be predicted by using the multiple regression. Each resource has relation between one to another, while information sources based on internal consistency test also mutual representing, where the role of actor also represented by hit counts: each web page may contain more than one name of actor, it forms co-occurrence and then relations for generating edges, and unit analysis in SNA. Thus, there are more than one relations between resources and SNA.

IV. CONCLUSION

In this study, we have presented an analysis for finding behaviour of resources of the social network. The resources of social networks - actor/vertex, relation/edge, Web/document, or connection/path - have different behaviour toward the growth of social networks based on seed. Analysis with statistic computation produces a relation between resources and SNA (based on the degree of vertices). More than one relation among actor, vertex, relation, edge, web, papers, and degree of actors. The future work will involve the extraction of a social network to describe the research collaboration.